# Properties of Physical Systems: Transient Singularities on Borders and Surface Transitive Zones

Mark E. Perel'man[*)]
*Racah Institute of Physics, Hebrew University, Jerusalem, Israel*

**Abstract**  Certain alternative properties of physical systems are describable by supports of arguments of response functions (e.g. light cone, borders of media) and expressed by projectors; corresponding equations of restraints lead to dispersion relations, theorems of counting, etc. As supports are measurable, their absolutely strict borders contradict the spirit of quantum theory and their quantum evolution leading to appearance of subtractions or certain needed flattening would be considered. "Flattening" of projectors introduce transitive zones that can be examined as a specification of adiabatic hypothesis or the Bogoliubov regulatory function in QED. For demonstration of their possibilities the phenomena of refraction and reflection of electromagnetic wave are considered; they show, in particular, the inevitable appearing of double electromagnetic layers on all surfaces that formerly were repeatedly postulated, etc. Quantum dynamics of projectors proves the neediness of subtractions that usually are artificially adding and express transient singularities and zones in squeezed forms.

## 1. Introduction

In the classical monograph of von Neumann [1] had been stated that qualitative (alternative) properties of physical systems can be described by projective operators (projectors) indicating their complete presence or complete absence. To such primordial properties of all systems must be attributed the causality, locality, mass-spectrality. However, particular features of definite objects/ devices such as proportions of solid media, spatial or temporal restrictions of parameters magnitudes, passband of frequencies or limits of work duration of technical devices and so on also can be added with the expanding of such description of the (characteristic) properties of specific objects.

For specification and improvement of this general statement the projectors of supports of certain response functions (range of variables, outside of which functions must be equal zero) can be selected, and the determination of these supports can be considered as the defining of properties. For properties of such type it leads to the establishment of certain features or classes of allowable response functions $f(\xi)$ describable by the equations:

$$f(\xi) = P(\xi) f(\xi), \qquad (1.1)$$

where $\xi$ represents a set of variables, e.g. (t, **r**) and/or (E, **p**), $P(\xi)$ is the projector of predefined support. These expressions, which we named "the equations of constraints", demonstrate that $f(\xi) = 0$, if $\xi \notin \operatorname{supp} P(\xi)$, and remains absolutely arbitrary in all the rest (possible singularities on borders of supports will be considered below).

Allowable Fourier convolutions of equations of constraints (1.1) lead to so called dispersion relations or spectral representations in the case of infinite supports and can be expressed as the theorems of counting in the case of finite (usually symmetrized) supports. The analytic description of properties, i.e. the transition from a set of properties to algebra of their projectors

[*)]. E-mails: mark_perelman@mail.ru; m.e.perelman@gmail.com .

and then to integral relations, was initially presented in the article [2], the general approach and some its results and applications are described in [3] (symmetries of some groups of transformations lead to the equations of constraints in the weak topology, which are not considered here.)

Nevertheless, if the magnitude of support is measurable, the absolute strictness/ clearness of borders definition can be in conflict with a spirit of quantum approach (note that certain uncertainty is inevitable in classics also, e.g. [4]).

Close problems were discussed, for the first time, as far as I know, by Stueckelberg [5]: he had proposed the introduction of diffuse temporal-spatial boundaries rather than strict and sharp ones at the course of QED calculations (more concretely at the photon emission, where this diffuseness implicitly meant the duration of process). This idea became the basic one in the Bogoliubov theory, but instead of diffuseness he had introduced the special covariant function of intensity of interactions and their switching-on and -off at the course of intermediate stages of calculations, i.e. the gradual increasing of interaction intensity [6]. Such approaches can be presented as the further development of the adiabatic hypothesis of Ehrenfest in quantum theory.

Some generalizations and/or specifications of this representation seem possible:

1/. Both themselves the adiabatic hypothesis and its generalization by Bogoliubov can be considered by the theory of interactions duration: they really correspond with duration of elementary acts [7];

2/. Operations of interaction switching on and off can be applied not only to time, but also to spatial variables, to definite spatial layers or zones, to transitions between them;

3/. Introduction of a "diffusion" zone can be not an intermediate or even only a formal operation, but to get in some cases a quite definite physical sense as the determination of the intermediate, transient formation between two media, two regimes of work and so on;

4/. Some analogs of subtractions, which correspond to this limiting procedure, can be achieved at consideration of canonical evolution of projectors that can be comparable to examination of evolution of projectors of rays $P(\omega) = |out\rangle \langle in|$ in the Hilbert space [8].

The necessity of some smoothing, of flattening of the strict restrictions or of existence of some intermediate zone can be shown on rather simple, but well known example, on the phenomenon of light reflection from a surface, where it may be described as an inevitable appearance of transient layers on borders. The impossibility of consideration of light reflection directly on geometrical plane, i.e. the existence of optically strange formation on surface, was marked already by Newton at the description of total internal reflection [†]. The phenomenon of frustrated total internal reflection (FTIR) was repeatedly rediscovered (e.g. [9]), and it became evident that this effect is not unique one: the processes of reflection and refraction of light fluxes (pulses) must be accompanied by certain so called nonspecular effects (e.g. [10] and references therein).

For explaining its and some other deviations from the Fresnel laws, based on determination of strict geometrical borders, Drude [11, 12] had supposed a century ago the existence of intermediate optical layers on all media surfaces. Analogical models were suggested by several authors (e.g. [13]), but these approaches were not successful, since the origin of such stable transitive layers had not been elucidated.

---

[†]. I. Newton in the immortal "Philosophiae Naturalis Principia Mathematica" (I - XIV, Theorem L) had described the flattened trajectory of the totally reflected rays. Such trajectory must be attributed, of course, to any mechanically reflecting body under taking into account the inevitable deformation of reflection surface. The problem 20 in his "Optics" supposes a gradual diminution of media density toward their surface.

On the other hand we had shown that these zones must exist as the dynamical formation, i.e. must be appearing and evolving under influence of interactions of incident waves with surface (factually as induced surface polarization) and fastly decaying after their cessation [14].

Formerly the greater attention to such superficial phenomena preferentially in metals was connected with the skin-effect, the specific field penetration into the depth of conducting media. Now with development of so-called near field optics and, more widely, in connection with problems of nanotechnology, the consideration of features of surface electromagnetic fields in all substances becomes more and more desirable.

But if the appearance of some intermediate, transitive zones or corresponding singularities can be of a similar origin, the question of their general description becomes especially interesting. And it seems very desirable to connect them with a general set of properties of physical systems, e.g. with projectors. It could mean that to many relations, expressed via projectors of general properties of systems, should be added or out of them should be extracted terms corresponding to especial border structures or to initial and final steps of devices work and of processes' courses. Among them can be, for example, macroscopic objects (e.g. reactances of electric circuits, diffracted rays, appearing double electric or magnetic layers on borders of media, transient zone of density near surfaces of solids, etc.) and also some phenomena of the quantum field theory. All these features are connected with inertial features, with (microscopic) structure of surfaces and/or can be directly explained by the (quantum) structure of considered systems, including vacuum. The consideration of all these particular models and theories can be facilitated by, if a developing of general phenomenological theory will be possible.

The most simple and evident way to such theory consists in an introduction of "extended" or "flattened" projectors that must be converging to the usual form of unit Heaviside steps on any stage of calculations [15]. Parameters of flattening can be connected with uncertainty values, with predefined durations of processes, with models or particular theories. Mathematical substantiation of such opportunities is given in the Section 2.

The offered scheme seems artificial and can be justified only a posteriori, via comparison of its applications with the known physical phenomena. With this aim in the Section 3 the sufficiently general phenomena are considered: reflection and refraction of electromagnetic wave on flat dielectric surface. It leads to appearance of the double electric layers (DELs), i.e. of dynamical transient zones.

In the Section 4 another approach to the problem is examined: the quantum evolution of supports as measurable magnitudes. This approach leads to results that are close in certain aspects to the results of flattening of projectors.

In the Section 5 certain problems of field theory, which begin from comparison of the Bogoliubov method with flattening projectors, are considered. The possibility of flattening of covariant field propagators with taking into account their quantum evolution is briefly described. The results are summed in the conclusions with mentioning of certain perspectives for further investigations.

## 2. Flattening of projectors

In the absence of detailed information about examinable process or system it is expedient to consider some limits of needed restrictions with possibly flattened boundaries. On subsequent stages such revealed features would be analyzed, compared with known particular solutions and so on. To such possibilities we shall go via a reexamination of the usual determination of supports or their projectors.

The Heaviside unit step operators are determined via the expansion of unity:

$$\theta(x) + \theta(-x) = 1, \tag{2.1}$$

where $\theta(|x|) = 1$, $\theta(-|x|) = 0$ and they are not determined at x = 0: it is not essential when (2.1) is employed under the integral sign, etc.

But if a magnitude of considered functions in the singular point x = 0 can play a significant role, the expansion of unity on three terms would be considered[‡]:

$$\vartheta(x) + \varsigma(x) + \vartheta(-x) = 1, \tag{2.2}$$

here $\vartheta(x)$ is the projector of the *open* interval $(0,\infty)$; the point-type projector $\varsigma(x) = 1$ under x = 0 and $\varsigma(x) = 0$ under x $\neq$ 0 [§]. Therefore

$$\vartheta^2(x) = \vartheta(x); \quad \vartheta(x)\vartheta(-x) = 0; \quad \varsigma(x)\vartheta(\pm x) = 0; \quad \varsigma^2(x) = \varsigma(x). \tag{2.2'}$$

The extremely significant for all our constructions is the functional representation of this point-type projector [16]:

$$\varsigma(x) = \sum a_n \delta^{(n)}(x). \tag{2.3}$$

The number of coefficients in (2.3) and their values, i.e. the needed subtractions, must be determined by particular theories and models

Let's consider possibility to express these projectors via flattened functions satisfying the basic equality (2.2) and supposing the possibility of subsequent limiting procedure:

$$\vartheta(x \mid a) + \varsigma_{a,b}(x) + \vartheta(-x \mid b) = 1, \tag{2.4}$$

Elements of (2.4) can be considered as "flattened projectors" (f-projectors) that at $a, b \to 0$ uniformly approach to the Heaviside units or zero. For this aim we can proceed with the evident definition of projector as

$$\theta(x) = \int_0^\infty dt \delta(x-t).$$

In accordance with this definition f-projectors can be constructed as limits of integrals over δ-sequences, conforming to the condition (2.4) at each step:

$$\vartheta(x \mid a) = \int_a^\infty d\xi \delta(x - \xi, a) = \int_{-\infty}^\infty d\xi \vartheta(x - \xi + a)\delta(\xi, a);$$

$$\varsigma_{a,b}(x) = 1 - \vartheta(-x \mid b) - \vartheta(x \mid a). \tag{2.5}$$

The most useful Cauchy-Lorentz and Gauss δ-sequences

$$\delta_1(x \mid a) = \frac{1}{\pi} \cdot \frac{a}{x^2 + a^2}; \tag{2.6}$$

$$\delta_2(x \mid a) = \frac{1}{a\sqrt{\pi}} \exp(-x^2 / a^2) \tag{2.6'}$$

with $a \to 0+$ lead to the f-projectors:

$$\vartheta_1(x \mid a) = \frac{1}{2} - \frac{1}{\pi} \tan^{-1}(1 - x/a) ; \tag{2.7}$$

$$\vartheta_2(x \mid a) = \frac{1}{2}\{1 - erf(x/a - 1)\}. \tag{2.7'}$$

The first of them is intuitively more transparent as $\vartheta_1(0 \mid a) = 1/4$ and, correspondingly, $\varsigma_{a,b}(0) = 1/2$. The values of second one at zero argument are lesser evident.

Notice such formal representations for projector derivatives:

$$\frac{d}{dx}P(x \mid a) = \delta(x \mid a)P(x \mid a) \quad \text{or} \quad \delta(x \mid a) = \frac{d}{dx}\ln P(x \mid a) \tag{2.8}$$

that allows a consideration of f-projectors via equations on self values.

The Fourier transforms of f-projectors,

$$\tilde{\vartheta}_1(k \mid a) = \delta_+(k)\exp\bigl(ika - a|k|\bigr); \tag{2.9}$$

$$\tilde{\vartheta}_2(k \mid a) = \delta_+(k)\exp\bigl(ika - a^2 k^2 / 4\bigr), \tag{2.9'}$$

show their shift on the distance $a$ and the appearing of additional waves at $a \neq 0$ connected with surfaces (projector's) flattening and rapidly damped outside the layer.

Notice that on the one hand the whole of written above can be reduced to more simple form at joining of the point projector with one of the "step" f-projectors, i.e. at consideration of (2.1) instead of (2.2). On the other hand intermediate zone can be divided onto parts with independent or consecutive limiting procedures.

Let us construct now the modified f-Hilbert transforms, corresponding to f-projectors. As example for such construction the isotropic medium in the down half space with dielectric susceptibility $\varepsilon(\omega, \vec{r})$ can be considered. If the upper half plane is a vacuum, then $f(\omega, \vec{r}) = \varepsilon(\omega, \vec{r}) - 1$ must be zero for $z > 0$ that can be expressed in the form of constraint equation:

$$\theta(z)f(\omega, \vec{r}) = 0. \tag{2.10}$$

Such equations are equivalent, for functions of some classes of integration, to the Hilbert transforms [17]. In the theory of f-operators the equations of constrains would be rewritten, in accordance with (2.4), as

$$\bigl[\vartheta(z \mid a) + \tfrac{1}{2}\varsigma_a(z)\bigr]f(\omega, \vec{r}) = 0. \tag{2.11}$$

In the f-representation with (2.9) the equation (2.11) is represented after the Fourier transformation over z in the form of f-Hilbert relations:

$$\tilde{f}(\omega, k) = \frac{1}{\pi i} \int_{-\infty}^{\infty} dq \frac{\tilde{f}(\omega, q)}{q - k} \exp\bigl(ia(q - k) - a|q - k|\bigr), \tag{2.12}$$

where the principal value of integral is meant.

This representation evidently leads to the dispersion relations over momenta in the form of modified or refined Kramers-Kronig relations. It shows, for example, that the maximal deviations against the usual Hilbert transforms must be observable for big wave numbers, e.g. for the fine structure of scatterers. In particular, the absorption on such frequencies will be diminished, which means a remoter propagation of evanescent waves, etc. It diminishes the role of $q$, which are far from $k$ and, for example, at formation of $\operatorname{Re}\widetilde{f}_a(k)$ effectively $\operatorname{Im}\widetilde{f}_a(k\pm q)$ with $q\le 1/a$ only plays role.

Decomposition of exponents into series leads to the Kramers-Kronig type relations with additional (subtraction) terms, which usually would be phenomenologically entered. The method of f-projectors allows the determining of their kind, the dependence on a surface structure or on the initial conditions and so on.

The first terms of such decomposition lead to the expression:

$$\widetilde{f}(k) = \frac{1}{\pi i}\int\limits_{-\infty}^{\infty}dq\frac{\widetilde{f}(q)}{q-k} + \frac{a}{\pi}\int\limits_{-\infty}^{\infty}dq\widetilde{f}(q)\big(1+i\operatorname{sgn}(q-k)\big) + \dots \qquad (2.13)$$

For the dielectric susceptibility $\widetilde{f}(k) \to \varepsilon(\omega)-1 = \big(\varepsilon_1(\omega)-1\big)+i\varepsilon_2(\omega)$ and with taking into account the oddness of its imaginary part it gives that

$$\varepsilon_1(\omega)-1 = \frac{2}{\pi}\int\limits_{0}^{\infty}d\eta\frac{\eta\varepsilon_2(\eta)}{\eta^2-\omega^2} + \frac{2a}{\pi}\int\limits_{0}^{\infty}d\eta\varepsilon_2(\eta)\,, \qquad (2.14)$$

$$\varepsilon_2(\omega) = -\frac{2\omega}{\pi}\int\limits_{0}^{\infty}d\eta\frac{\varepsilon_1(\eta)}{\eta^2-\omega^2} - \frac{2a}{\pi}\int\limits_{0}^{\infty}d\eta\theta(\omega-\eta)\varepsilon_1(\eta)\,, \qquad (2.14')$$

i.e. both parts of dielectric susceptibility are renormalizable by surface states. A depth of transient zone (time) is determined by the parameter $a$ and can represent, in particular, an evanescent wave in the alternative field.

The parameter $a$ can be expressed via appropriate magnitudes such as uncertainty values, thermodynamic or microscopic quantities.

## 3. Electromagnetic waves

Let's consider an electromagnetic wave that falls from free space (z>0) onto medium (z<0). In the method of Bremmer ([18], Appendix 6) each component of fields $\vec{E},\vec{D},\vec{B},\vec{H},\vec{j}$ or $\vec{E},\vec{B},\vec{D},\vec{j}$ at another description of fields in substance [19] is expressed at the border via sum of other components multiplied on the suitable unit step Heaviside functions:

$$V(t,\vec{r}) = (V_I + V_R)\theta(z) + V_T\theta(-z)\,. \qquad (3.1)$$

The quantum evolution considered below will lead to the expansion of projectors and appearance of singularities, joining of which can be represented as a flattening of supports of projectors. It would complicate the problem as the special detachment of parts related to an appearing stratum will be needed. Instead of it we can *suppose* the primary flattening of supports via transition from the very beginning to the f-formalism and with the introduction of intermediate (near-surface) waves:

$$V(t,\vec{r}) = (V_I + V_R)\vartheta(z\,|\,a) + V_T\vartheta(-z\,|\,b) + V_0\varsigma_{a,b}(z)\,, \qquad (3.2)$$

where the subscripts *I, R, T, 0* are referring, correspondingly, to the initial, reflected, transmitted and intermediate, i.e. near fields and currents, $a$ and $b$ are depths of an intermediate layer (its spreading in both half-spaces). At $a = b = 0$ this decomposition transfers into (3.1) with neglecting of near field effects. Note that at the asymmetric approach to Maxwell equations in substance the opportunity of occurrence of singular currents along an inter-surface is considered for the problems of light reflection without obvious introduction of an intermediate layer. The used boundary conditions are similar to the limiting form of (3.2) [20].

Notes that the form (3.2) is maintained in quadratures, i.e. for the energy and momenta flows:

$$|\vec{V}(t,\vec{r})|^2 = |\vec{V}_I + \vec{V}_R|^2 \ \vartheta(z\,|\,a) + |\vec{V}_T|^2 \ \vartheta(-z\,|\,b) + |\vec{V}_0|^2 \ \varsigma_{a,b}(z)\,. \qquad (3.3)$$

For consideration of these fields the general form of equations

$$\vec{\nabla}\cdot\vec{V} = 4\pi(\rho + \rho^{(fict)})\,; \qquad (3.4)$$

$$\vec{\nabla}\times\vec{V} = \frac{i\omega}{c}\vec{W} + \frac{4\pi}{c}(\vec{j} + \vec{j}^{(fict)})\,, \qquad (3.5)$$

with additional, fictive or virtual charges and currents must be used.

Let's describe the TE wave in the (x, z) plane that is entering into an electrically neutral and optically passive substance. Since all fields, except $V_0$, satisfy the equation $\vec{\nabla}\vec{V} = 4\pi\rho$ ($\rho_i \to 0$ for metals) and $\vec{\nabla}\vec{\varsigma}(z\,|\,a,b) = d\varsigma\,/\,dz$, the operations of divergence of (3.2) with $V_i \to D_i$ leads to the relation:

$$\vec{\nabla}\cdot\vec{D}(z) = D_{0z}\bigl(\delta(z,a) - \delta(z,-b)\bigr) + 4\pi\rho_0^{(e)}(z)\varsigma_{a,b}(z) = 0\,. \qquad (3.6)$$

Hence the intermediate layer represents an oscillating double electric layer of strength $D_{0z}$ with induced charges of density $\rho^{(e)}$.

For the case of TM wave in the incident plane, we apply (3.2) to the magnetic field, $V_i \to B_i$, with taking into accounts that $\vec{\nabla}\vec{V} = 0$ for all fields except $V_0$:

$$\vec{\nabla}\cdot\vec{B}(z) = B_{0z}\bigl(\delta(z,a') - \delta(z,-b')\bigr) + 4\pi\rho_0^{(m)}(z)\varsigma_{a',b'}(z) = 0\,. \qquad (3.7)$$

where $\rho^{(m)}$ is the density of "magnetic charges" inducing in a near-surface zone by transition effects. In diamagnetics, e.g. in Ag, this field increases the absolute value of existing $\mu < 0$.

The similarity of last expression with (3.6) demonstrates not only a resemblance of both considered results, but also the difference of their possibilities at description of higher moments, etc.

These expressions can be rewritten as

$$D_{0z}\kappa(a,b) = 4\pi\rho_0^{(e)}\,; \qquad B_{0z}\kappa(a,b) = 4\pi\rho_0^{(m)}\,, \qquad (3.8)$$

where

$$\kappa(a,b) = \frac{d}{dz}\ln\varsigma_{a,b}(z) \qquad \text{or} \qquad \varsigma_{a,b}(z) = \exp\bigl(z\kappa(a,b)\bigr)\,. \qquad (3.9)$$

Physically it can be *supposed* that $1/\kappa$ would be of the order of generalized "skin layer" thickness and $\varsigma_{a,b}$ can be considered as the response function of this stratum.

If the falling wave is of the frequency $\omega$ and is incident under the angle $\alpha$, both parameters can be expressed via the uncertainty principle:

$$a = \frac{c}{\varepsilon_1\omega\cos\alpha}\,; \qquad b = a\sqrt{\varepsilon_1\,/\,\varepsilon_2}\,. \qquad (3.10)$$

Such simplest estimation seems completely adequate for the phenomenon of FTIR [13] and can be improved for more realistic or complicated cases with taking into account different parameters of medium.

The substitutions of (3.2) into other Maxwell equations describe the evanescent, oppositely directed currents in this layer. Thus, oscillating "dipoles" and "currents" on frequencies of incident fields absorb falling radiation and emit reflected and refracted waves.

Note that these equations lead to a wave equation with complicated right side depending on fictive charges, currents and material functions:

$$\left(\nabla^2 - c^{-2}\partial_t^2\right)\vec{V}_0 = f(\varepsilon, \mu; \vec{j}, \rho)_{fict} . \tag{3.10}$$

Their considerations has sense at introducing definite particular models only and can be here omitted (compare [21]).

The consideration of acoustic waves is very close to all above and can be also omitted.

## 4. Quantum evolution of projectors

Alternative properties of quantum system at one moment of time can be described by expressing their projectors (supports) in terms of space-time variables. A given projection can be re-described in terms of the coordinates at any time as in the Heisenberg picture these coordinates evolve in time. Thus at any time it on-going represents some alternative property (cf. [8]).

The time dependence of each operator, including projectors, is expressed in the Heisenberg representation as the shift over time (all at c= $\hbar$ =1):

$$P(t + \tau, \vec{r}) = e^{i\tau\hat{H}} P(t, \vec{r}) e^{-i\tau\hat{H}} , \tag{4.1}$$

which leads to the equation of evolution of this projector:

$$\frac{\partial}{i\partial t} P(t, \vec{r}) = \left[\hat{H}, P(t, \vec{r})\right], \tag{4.2}$$

the state $\Psi$ is unchanging in time.

The general expression (4.1) is uncovered as

$$P(t + \tau) = P(t) + [i\tau\hat{H}, P(t)] + \frac{1}{2}[i\tau\hat{H}, [i\tau\hat{H}, P(t)]] + \dots \rightarrow$$
$$\rightarrow P(t) + \sum_1^\infty a_n(t, \tau)(\partial/\partial t)^n P(t) \tag{4.3}$$

Thus, the evolving of projector adds to it a series of $\delta$-functions and their derivatives, arguments of which correspond to the borders of initial support. (The alterations of spatial variables are, for simplicity, omitted.)

Hence this expression can be evidently comparing with (2.3), i.e. with the implicit introduction of the zero-point projector. Indeed just this circumstance with physically reasonable flattening of singular functions can be examined as a substantiation of the scheme offered in the Section 2. So it can be stated that the flattening of projectors does not contradict general laws of their quantum evolution and even represents some their approximation.

Let us consider the simplest example: a device that is working during strictly defined temporal interval:

$$f(t) = \theta(T^2 - t^2) f(t) . \tag{4.4}$$

Its response function is expressed via the Fourier transformation as the Duhamel integral:

$$f(\omega) = \frac{1}{\pi} \int d\eta \frac{\sin(\omega - \eta)T}{(\omega - \eta)} f(\eta) \tag{4.5}$$

(at summing instead integration it leads to the Shannon theorem of counting).

At temporal shift of variable in projector, $t \rightarrow t + \tau$, the corresponding shift takes place in the integral representation. Let's consider this shift via (4.3). If the Hamiltonian $\hat{H} \rightarrow \partial / i\partial t$, then

$$[i\tau\hat{H}, \theta(T^2 - t^2)] = \tau \, \text{sgn}(T)[\delta(T - t) - \delta(T + t)], \tag{4.6}$$

i.e. represents the double (temporal) layer. If the responses function has only the restricted number of derivatives, it will show an existence of definite number of subtractions, some of additional terms can correspond to features of device switching on and off.

In the simple electric circuits with constant elements only two subtractions that reflect the conductivities and inductees elements are needed, in more complicated circuits additional terms will correspond to their alterations. The magnitudes of $\tau$ correspond to temporal parameters of elements or to their uncertainty values.

Let's examine other possibilities of description of systems evolution over shifts analogical to (4.1), but with other variables. The equation (4.2) is directly extended into the equation of motion in the Heisenberg representation with the operator of 4-momentum:

$$\frac{\partial}{i\partial x_\mu} P(x) = [p_\mu, P(x)], \tag{4.7}$$

$x = (t, \vec{r})$, and correspondingly the space shifts can be considered.

So by the complete analogy with (4.4) the 1D slit of 2A width in $x$ direction in the completely absorbing barrier can be described by the equation of restriction:

$$f(t, \vec{r}) = P(t, x) f(t, \vec{r}) \equiv \theta(A^2 - x^2) f(t, \vec{r}). \tag{4.8}$$

With $\hat{H} \rightarrow \hat{p}_x = i\partial / \partial x$ and $\tau \rightarrow \xi$ the analog of (4.3) requires calculations of commutators:

$$[i\xi\hat{p}_x, \theta(A^2 - x^2)] = \xi[\delta(A - x) - \delta(A + x)], \tag{4.9}$$

i.e. it describes a double electrical (magnetic) layer, DEL or DML, on boundaries of slit, higher derivatives can correspond to their alterations and to layers of higher types. These layers must oscillate with frequencies of falling waves, and oscillations of double layers lead to secondary radiation [22, 23]; such phenomena that can depend on features of screen and its borders, as far as I know, never were examined.

The border, for simplicity flat, between two media can be described analogically:

$$\varepsilon(z) = \varepsilon_1 \theta(z) + \varepsilon_2 \theta(-z) = \varepsilon_2 + (\varepsilon_1 - \varepsilon_2)\theta(z) \tag{4.10}$$

with constant $\varepsilon_n$ (instead a dielectric susceptibility another suitable material characteristics can be considered). The quantum evolution of the system leads to appearance of subtractions that depend on alterations of material parameters. This expression can be naturally generalized on the case of flattening of varying parameters:

$$\varepsilon(z) = \varepsilon_2 + \sum \Delta\varepsilon(z)(\theta(z_n) - \theta(z_{n-1})) \rightarrow \int d\zeta \cdot \varepsilon(\zeta)\delta(z - \zeta). \tag{4.10'}$$

For consideration of restrictions over admissible energy and/or momenta and corresponding projectors can be used the equation reciprocal to the Schrödinger equation:

$$i\frac{\partial}{\partial \omega} P(E, \vec{r}) = [\hat{\tau}, P(\omega, \vec{r})], \tag{4.11}$$

where instead of Hamiltonian the operator of interaction time duration is taken [7]. It leads to expansions analogical (4.3). Strictly denoted pass-band of devices can be considered analogously, by the substitutions $(T, t) \rightarrow (\Omega, \omega)$. They lead to analogs of (4.6) in the temporal representation.

Thus, *quantum evolution can expand the initial temporal (geometrical) domains of considered processes or interactions; which can induce the additional subtractions of response functions. It means that the strict initial restrictions can be or even should be broken or flattened in the course of quantum evolution.*

## 5. To field theory

Let us compare these constructions with certain problems of the field theory.

The Bogoliubov approach to the quantum field theory is based on introduction of the covariant functions of switching (or inclusion) of interaction g(x), that is inconcretized and only satisfies the requirements (in this Section we use the common 4D relativistic notation, $\hbar = c = 1$):

a. $g(x) \in [0, 1]$. This condition can be slightly generalized as $|g(x)| \in [0, 1]$;

b. General covariance;

c. Limiting conditions: $g(x) \to 0$ at $x_0 \to \pm\infty$;

d. $g(-x_\mu) = g(x_\mu)$, this condition follows in QED from the requirements of CPT invariance of the 4-potential $A_\mu$.

The f-projectors (2.7) completely correspond to these requirements with the substitution $x/a \to \gamma |n_\mu x_\mu|$, where $n_\mu$ is an unit time-like 4-vector and in particular the value $n_\mu \to \delta_{\mu,0}$ can be taken. The choice of parameter $\gamma$ depends on considered problems.

Thus there is not the significant divergence of offered approach with the canonical theory. Moreover, it can be suggested that it represents the *particular* realization of this theory.

The approach to evolution of projectors can be directly applied, for example, to the evolution of the restriction of relativistic causality,

$$f(t, \vec{r}) = \theta(-t)\theta(t^2 - r^2) f(t, \vec{r}) \equiv \theta(-t- |\vec{r}|) f(t, \vec{r}), \tag{5.1}$$

that can be considered as above, via (4.2) and/or (4.3). But, in addition, it can be expressed via Green functions of the reciprocal Klein-Gordon equation introduced in [2]:

$$(x^2 - s^2) f(x) = \delta(x) \quad \text{or} \quad \left(\partial_E^2 - \partial_p^2 + s^2\right) f(p) = 1. \tag{5.2}$$

Here $s$ can be attributed as the operator of 4-interval and the Green functions of (5.2) are presentable as

$$\nabla^{(ret)}(p, s) = -i(2\pi)^3 \int dx e^{ipx} \theta(-t) \delta(x^2 - s^2), \tag{5.3}$$

that is simply connected with the Green functions of Klein-Gordon equation:

$$\nabla^{(.)}(p, s \mid ret, adv; \pm) = \Delta_{(.)}(x \to p, m \to s \mid \mp; ret, adv). \tag{5.4}$$

The estimation and regularization of $\nabla$ functions exactly follows the theory of $\Delta$ functions with substitutions (5.4). The derivatives and products of these functions are expressed via them with parameters $s_i$, different values of 4-interval instead of different masses (we completely follow [6]) and therefore for regularization can be used the reciprocal Pauli-Villars method that leads to

$$reg\{\nabla(x, s)\} = \nabla(x, s) + \sum_1^n \alpha_i \nabla(x, s_i), \tag{5.5}$$

i.e. the regularization introduces processes with different velocities non equal $c$. It means appearance of different virtual particles at evolution process participating in propagation of initial particle, the phenomenon close to dressing.

Projector of relativistic causality in (5.1) can be now expressed as

$$P_{RC}(x) = \int_0^\infty ds \nabla^{(ret)}(x, s) \tag{5.6}$$

and its evolution can be expressed via evolution of the corresponding "reciprocal" Green function. Thus processes with different velocities extenuated surface of light cone, flattened it.

The restrictions of positivity of energy and mass-spectrality [2],

$$f(p) = P_{E,m}(p)f(p) \tag{5.7}$$

with the projector

$$P_{E,m} = \theta(E)\theta(p^2 - m^2) = (2\pi)^3 i \int_m^\infty d\mu \mu \Delta^{(+)}(p, \mu), \tag{5.8}$$

can be analyzed with the reciprocal equations of quantum dynamics [7] of the type (4.11) that contained the operator of interaction duration instead the Hamiltonian. It can lead to masses instability, etc. For this case the considered procedure of regularization can be applied without any modification.

**CONCLUSIONS**

Let us sum the results and mention certain possibilities of developed methods.

1. Certain properties of physical systems (in the von Neumann sense) are describable by projectors to which can be comparable definite supports of response functions. To these common properties can be added some constrained characteristics of devices and bodies. These possibilities lead to the dispersion relations, theorems of counting and so on.

2. Strict borders of these supports can contradict the spirit of quantum theory. As these supports are measurable, their borders may be smoothed, flattened, at least, in the sense of uncertainty principles.

3. This flattening can be considered as certain concretization of the adiabatic hypothesis and/or the Bogoliubov method of regularization of interaction's intensity.

4. Such smoothing can be introduced phenomenologically, via primary flattening of projectors of supports. The consistent theory of flattening of projectors is constructed. Their introduction implies subsequent comparison of parameters of these expanded, diffuse boundaries with partial theories, models, etc. Note that the offered approach can have a definite heuristic significance also.

5. This theory is applied to phenomena of reflection and refraction of electromagnetic waves on flat surface. The examination demonstrates the appearance of double electric or magnetic layers on optical boundaries, elements of such diffuse layers absorb falling wave and emit reflected and transmitted waves. These effective zones can execute the role of additional layers postulated by Drude and others for improving the Fresnel formulae.

6. It is shown that the possibility of an existence of double electric (magnetic) layers on surfaces of condensed media represents their general feature or even the necessity. In liquid near-surfaces their formation can be connected with the asymmetry of molecules [24] and their turning; near to solid surfaces they would be connected with holes formation and their movement [25]. The inevitable existence of such layers may have a special significance at processes of condensation and solidification (cf. [26]). Their radiation can be determinative at surf noises, for certain possibilities of prognosis of earthquakes [22, 23], etc. The existence of transient zones, as must be underlined, does not depend in itself on microscopic structure of substances.

7. The especial quantum approach via consideration of quantum temporal or spatial shifts of projectors leads to the appearance of singular terms on the borders of supports. Their consideration shows that they can represent squeezed forms of transient zones introduced via flattening of projectors. Thus their consideration justifies, *a posteriori*, procedures of introduction of flattening of projectors.

Let us briefly mention certain other perspectives of the offered approach.

Analogical consideration of acoustic waves would lead to revealing of a peculiar transient zone over density on borders of media.

The extremely interest, both from the applied point of view, and for substantiation of our theory represent the supervisions of Distler of long-range ordering through metal - nonmetal interfaces, i.e. a distant transfer of the information on structure of crystals (e.g., [27]). In these experiments the grown crystal had been completely capsulated by a nano-film, then it had been placed into parent solution with renewing of its growth: very strangely at first glance but the growing crystal repeats distinctive features of closed surfaces (defects, dislocations, steps [28]). These supervisions show that formed double layers are effective, i.e. their fields extend further geometrical borders of a body and they can regulate the crystallization process.

Among other possible applications can be mentioned out possibilities of calculations of optical apertures with taken into account the flattening of rays (cf. [29]).